# Multifunctional Two-dimensional van der Waals Janus Magnet Cr-based Dichalcogenide Halides

Yusheng Hou[1,#,*], Feng Xue[2,#], Liang Qiu[1], Zhe Wang[3] and Ruqian Wu[4,*]

[1] Guangdong Provincial Key Laboratory of Magnetoelectric Physics and Devices, Center for Neutron Science and Technology, School of Physics, Sun Yat-Sen University, Guangzhou, 510275, China

[2] State Key Laboratory of Low-Dimensional Quantum Physics, Department of Physics, Tsinghua University, Beijing 100084, China and Frontier Science Center for Quantum Information, Beijing 100084, China

[3] State Key Laboratory of Surface Physics and Key Laboratory for Computational Physical Sciences (MOE) and Department of Physics, Fudan University, Shanghai 200433, China

[4] Department of Physics and Astronomy, University of California, Irvine, CA 92697-4575, USA

# Contributed equally to this work

Corresponding authors (*): houysh@mail.sysu.edu.cn; wur@uci.edu

# Abstract

Two-dimensional van der Waals Janus materials and their heterostructures offer fertile platforms for designing fascinating functionalities. Here, by means of systematic first-principles studies on van der Waals Janus monolayer Cr-based dichalcogenide halides Cr$YX$ ($Y$=S, Se, Te; $X$=Cl, Br, I), we find that CrS$X$ ($X$=Cl, Br, I) are the very desirable high $T_C$ ferromagnetic semiconductors with an out-of-plane magnetization. Excitingly, by the benefit of the large magnetic moments on ligand $S^{2-}$ anions, the sought-after large-gap quantum anomalous Hall effect and sizable valley splitting can be achieved through the magnetic proximity effect in van der Waals heterostructures CrSBr/Bi$_2$Se$_3$/CrSBr and MoTe$_2$/CrSBr, respectively. Additionally, we show that large Dzyaloshinskii-Moriya interactions give rise to skyrmion states in CrTe$X$ ($X$=Cl, Br, I) under external magnetic fields. Our work reveals that two-dimensional Janus magnet Cr-based dichalcogenide halides have appealing multifunctionalities in the applications of topological electronic and valleytronic devices.

## INTRODUCTION

The discovery of ferromagnetism at a finite temperature in atomically thin van der Waals (vdW) monolayers (MLs) [1-3] has spurred a surge of experimental and theoretical interests in understanding the two-dimensional (2D) magnetism and furthermore in producing emergent functionalities in 2D vdW heterostructures[4-9]. Fundamentally, the well-known Mermin-Wagner theorem based on the isotropic Heisenberg exchange model [10] suggested the absence of either FM or antiferromagnetic (AFM) order in 2D systems at nonzero temperature, and hence the early findings of FM order in $CrI_3$ ML [1] and $Cr_2Ge_2Te_6$ bilayer [11] were rather surprising. Recent studies indicated that magnetic anisotropies play a major role in the establishment of their long-range magnetic ordering. The main advantage of these 2D magnetic vdW materials is their integrability with other vdW functional materials. Through the magnetic proximity effect, many spintronic, valleytronic and optoelectronic properties can be achieved in heterostructures by stacking diverse 2D vdW MLs or ultrathin films [5,7,8]. For example, the quantum anomalous Hall effect (QAHE) and axion insulator phase can be realized by integrating 2D FM semiconductors with thin films of topological insulators [8,12-15]. When transition metal dichalcogenide (TMDC) MLs are in contact with 2D FM semiconductors, a valley splitting can be produced for valleytronic and optoelectronic manipulations [7], as recently observed in $WSe_2/CrI_3$ [16]. Obviously, it is beneficial to have large spin polarization in the outer anions of 2D FM MLs to maximize the magnetic proximity effect.

Recently, Janus 2D materials have attracted increasing attention [17,18] since TMDC Janus ML MoSSe was successfully fabricated by controlling the reaction conditions in experiments [19]. Due to the feasibility of selecting a suitable pair of anions, Janus 2D materials add an extra dimension for the rational design of 2D vdW MLs with desired properties. For example, because the out-of-plane symmetry is broken, 2D Janus CrGe(Se,Te)$_3$, Cr$X$Te ($X$=S, Se) and manganese dichalcogenide MLs may possess large Dzyaloshinskii-Moriya interaction (DMI) and thus can host magnetic skyrmions [20-23]. Besides, Janus structure as electrodes in Li-ion batteries may have good performance due to its structural asymmetry [24]. Several theoretical studies [25-28] predicted that Janus MLs, such as VSSe [27], a Janus structure of the FM VSe$_2$ [2,29], have FM order up to room-temperature. Although 2D Janus magnets' own properties have been widely investigated, their heterostructures with other vdW functional materials, such as TMDC MLs and three-dimensional (3D) topological insulator thin films, remain underexplored [17,18]. Moreover, the formation of complex magnetic patterns in 2D heterostructures due to the competition among Heisenberg exchange, magnetic anisotropy, DMI and external magnetic field is rarely discussed.

In this work, we systematically study the electronic and magnetic properties of vdW Janus ML Cr-based dichalcogenide halides Cr$YX$ ($Y$=S, Se, Te; $X$=Cl, Br, I) using first-principles calculations. This is partially inspired by the experimental observation of high-temperature (~200 K) ferromagnetism in 1$T$-CrTe$_2$ ML[30-34]. We identify that CrS$X$ ($X$=Cl, Br, I) are the very attractive out-of-plane FM semiconductors with high Curie temperatures (~176 K) and large magnetic moments on $S^{2-}$ anions. Excitingly, through the magnetic proximity effect, the sought-after large-gap QAHE can be achieved in CrSBr/Bi$_2$Se$_3$/CrSBr. For the same reason, a sizable valley splitting of 37.9 meV, corresponding to a magnetic field of 379 Tesla[35], can be generated in MoTe$_2$/CrSBr. Furthermore, we find that large DMI leads to magnetic skyrmion states in CrTe$X$ ($X$=Cl, Br, I) under an appropriate external magnetic field by virtue of Monte Carlo (MC) simulations. Our work highlights the remarkable multifunctionalities of these 2D vdW Janus ML magnets that are promising for applications in the next-generation topotronic and valleytronic devices.

## RESULTS

### Electronic and magnetic properties of Cr$YX$ ML

Fig. 1a shows the crystal structure of Janus ML Cr$YX$ ($Y$=S, Se, Te; $X$=Cl, Br, I). Magnetic Cr$^{3+}$ cations form a triangular lattice and are sandwiched by $Y^{2-}$ and $X^-$ anions. Similar to CrI$_3$ and Cr$_2$Ge$_2$Te$_6$ MLs [1,11], Cr$^{3+}$ cations in Cr$YX$ are surrounded by the distorted edge-shared octahedrons formed by $Y^{2-}$ and $X^-$ anions, which implies FM nearest neighbor (NN) exchange interactions. As a direct result of different sizes of $Y^{2-}$ and $X^-$ anions, CrSCl and CrTeI have the smallest and largest in-plane lattice constants (Supplementary Table 1), respectively. Since no imaginary frequency is found in the calculated phonon spectra of all Cr$YX$ MLs (Supplementary Fig. 1), they should be dynamically stable.

To explore the magnetic ground states of Cr$YX$ MLs, we consider a spin Hamiltonian consisting of Heisenberg exchange interactions, DMI and single ion anisotropy (SIA).

This spin Hamiltonian is in the form of [21,22]

$$H = J_1 \sum_{\langle ij \rangle} \mathbf{S}_i \cdot \mathbf{S}_j + J_2 \sum_{\langle\langle ij \rangle\rangle} \mathbf{S}_i \cdot \mathbf{S}_j + J_3 \sum_{\langle\langle\langle ij \rangle\rangle\rangle} \mathbf{S}_i \cdot \mathbf{S}_j + \sum_{\langle ij \rangle} \mathbf{D}_{ij} \cdot \left( \mathbf{S}_i \times \mathbf{S}_j \right) - A \sum_i \left( S_i^z \right)^2 \qquad (1).$$

Here, $\mathbf{S}_i$ is the normalized spin vector at site $i$; $J_1$, $J_2$ and $J_3$ are NN, second-NN and

third-NN Heisenberg exchange parameters, respectively; $\mathbf{D}_{ij}=(D_x, D_y, D_z)$ is the NN

DMI vector and $A$ is the SIA parameter. The Heisenberg exchange parameters $J_i$ ($i$=1,

2, 3), NN DM vector $\mathbf{D}_1$ and SIA parameter $A$ from DFT calculations are tabulated in

Supplementary Table 2. As expected, the NN $J_1$ is FM and dominates over the second-

NN $J_2$ and third-NN $J_3$ for all Cr$YX$ (Fig. 1b). Since DMI is directly related to the spin-orbit coupling (SOC) [36], the magnitude of the NN DMI vector, $|\mathbf{D}_1| = \sqrt{D_x^2 + D_y^2 + D_z^2}$, increases when $Y$ ($X$) goes from S (Cl) to Te (I), as shown in Fig. 1c. It is interesting that $\left|\mathbf{D}_{1,//}\right|/\left|J_1\right|$ ($\mathbf{D}_{1,//}$, the inplane component of $\mathbf{D}_1$) has a similar trend to $|\mathbf{D}_1|$ when $Y$

($X$) varies (Fig. 1c). Particularly, $\left|\mathbf{D}_{1,//}\right|/\left|J_1\right|$ of CrTe$X$ ($X$=Cl, Br, I) is in the typical

range of 0.1-0.2 that is known to generate magnetic skyrmions [37]. Lastly, CrS$X$ ($X$=Cl,

Br, I) and CrSeI have an out-of-plane SIA while the other five of Cr$YX$ have an in-plane

SIA (Fig. 1d). Note that the present results of NN FM Heisenberg exchange interactions and in-plane SIAs of CrSeBr and CrTeI Janus MLs are consistent with those obtained in a previous theoretical study [38].

Owing to the competition between Heisenberg exchange interactions, DMI and SIA,

MC simulations reveal that Cr$YX$ exhibits very rich magnetic ground configurations

(Fig. 1e). It has been reported that magnetic ground state of 2D magnets is mainly determined by a critical dimensionless factor $|A| K / \mathbf{D}_{//}^2$ ($K$, the stiffness parameter originating from FM Heisenberg exchange interactions)[39]. A large $|A| K / \mathbf{D}_{//}^2$ yields a FM ground state, while a small one results in a spiral ground state [39]. First, CrS$Y$ ($Y$=Cl, Br, I) MLs have small DMI, out-of-plane SIA and hence an out-of-plane FM ground state with a Curie temperature up to 176 K (Supplementary Fig. 2). CrSeCl also has a FM ground state, but its magnetization is in plane due to its negative SIA. Second, CrSeBr and CrSeI have medium DMI, small SIA, and hence small $|A| K / \mathbf{D}_{//}^2$. Therefore, they have a spin spiral ground state with a large periodic length. Finally, CrTe$X$ ($X$=Cl, Br, I) MLs have both large DMI and SIA. Their magnetic ground states have wormlike domains, similar to what was found in the previous studies of 2D Janus manganese dichalcogenides [20].

Considering the needed semiconducting nature of vdW ferromagnets for engineering emergent physical properties via the magnetic proximity effect in heterostructures [7,13,16], we first analyze the electronic properties of CrS$X$ ($X$=Cl, Br, I) MLs which have a FM ground state and out-of-plane magnetic anisotropy. Band structures in Figs. 2a-2c show that all CrS$X$ MLs are indirect-gap semiconductors. As the electronegativity weakens from Cl to Br and to I, the band gaps of CrS$X$ decrease from 2.17 eV (CrSCl) to 1.77 eV (CrSBr) and to 0.81 eV (CrSI). The curves of density of state (Supplementary Fig. 3) suggest that the conduction bands mainly come from the $d$ states of Cr while the valence bands have mixtures of Cr, S and $X$ states. Besides, the $d$-$p$ hybridization between Cr and S atoms is stronger than that between Cr and $X$ atoms near the Fermi level.

As the magnetic ions are mostly covered by nonmagnetic anions in vdW FM semiconductors, the spin polarization of the outer layer is typically very small. The magnetic proximity effect is hence weak in most vdW heterostructures. Interestingly, the magnetic moments on $S^{2-}$ anion are about 0.14 $\mu_B$/S in CrS$X$ MLs, the largest for 2D FM semiconductors reported so far (Fig. 3c) [1,11,40-45]. Especially, this magnetic moment is much larger than that of $I^{2-}$ anions in CrI$_3$ [7,13,16,46-49], suggesting that CrS$X$ may have a significant magnetic proximity effect on other vdW functional materials.

**Large-gap QAHE in CrSBr/Bi$_2$Se$_3$/CrSBr vdW heterostructure**

We first examine the effect of CrS$X$ MLs on magnetizing topological surface states (TSSs) of 3D topological insulators for the realization of QAHE [8]. We construct an vdW heterostructure with two CrSBr MLs and a six quintuple-layer Bi$_2$Se$_3$ (BS) thin film (CrSBr/BS/CrSBr). The calculated binding energies of different stacking configurations suggest that Cr$^{3+}$ cations of CrSBr prefer to align with Bi and the $S^{2-}$ side contacts BS thin film (Fig. 3a and more details in Supplementary Fig. 4). Our *ab initio* molecular dynamics simulations show that this CrSBr/BS/CrSBr heterostructure is thermodynamically stable (Supplementary Fig. 5). The induced spin polarization on BS penetrates through the film (Fig. 3b), with a strength twice as large as that in CrI$_3$/BS/CrI$_3$ [13]. Fig. 3c shows the band structure of CrSBr/BS/CrSBr when the magnetizations of top and bottom CrSBr MLs are ferromagnetically ordered. As a result of the strong magnetization in CrSBr, the system has a large band gap up to 19 meV at the $\Gamma$ point, indicating its efficient magnetic proximity effect on the TSSs of BS thin film. By examining the spin components of the four bands near the Fermi level (Fig. 3d), we find that the two bands below (above) the Fermi level have the same spin-down (spin-up) components. These features clearly suggest the TSSs of the BS thin film are strongly magnetized by CrSBr.

To investigate the topological property of CrSBr/BS/CrSBr, we fit its band structure using an effective four-band model. With the bases of $\left\{ \left| t,\uparrow \right\rangle, \left| t,\downarrow \right\rangle, \left| b,\uparrow \right\rangle, \left| b,\downarrow \right\rangle \right\}$, the model Hamiltonian of inversion symmetric heterostructures is written as [50]

$$H\left(k_x,k_y\right)=Ak^2+\begin{bmatrix} v_F\left(k_y\sigma_x-k_x\sigma_y\right) & M_k\sigma_0 \\ M_k\sigma_0 & -v_F\left(k_y\sigma_x-k_x\sigma_y\right) \end{bmatrix}+\begin{bmatrix} \Delta\sigma_z & 0 \\ 0 & \Delta\sigma_z \end{bmatrix} \quad (2) \cdot$$

In Eq. (2), $\uparrow$ ($\downarrow$) denotes spin-up (spin-down) states; $v_F$, $k^2=k_x^2+k_y^2$ and $\sigma_{x,y,z}$ are the Fermi velocity, in-plane wave vector and Pauli matrices, respectively. The coupling between the top and bottom TSSs of BS thin film is described by $M_k=\Delta_h+Bk^2$ and $\Delta_h$ is the coupling induced gap. By fitting the DFT calculated band structure with Eq. (2) (Supplementary Fig. 6), we obtain $\Delta_h$ and $\Delta$ are 1.0 and 10.4 meV, respectively. According to the general rule proposed in Ref. 13, CrSBr/BS/CrSBr is a Chern insulator with Chern number $C_N$=1. This topological nature is further confirmed by the presence of one chiral edge state that connects the valence and conduction bands in the band structure of the one-dimensional CrSBr/BS/CrSBr nanoribbon (Fig. 3e). Taking together $T_C$ (176 K) of the FM semiconductor CrSBr ML and the large nontrivial band gap (19 meV, corresponding to 220 K), it is conceivable that QAHE with a high temperature can be achieved in CrSBr/BS/CrSBr.

**Sizable valley splitting in CrSBr/MoTe₂ vdW heterostructure**

We further explore the valley splitting of TMDC MLs in contact with CrSBr. To this end, we study the vdW heterostructure of CrSBr and MoTe₂ MLs (CrSBr/MoTe₂). Again, the calculated binding energies (Supplementary Fig. 7) indicate that the $S^{2-}$ side of CrSBr binds to MoTe₂ (Fig. 4a). $Cr^{3+}$ cations and $S^{2-}$ anions sit on the top of $Mo^{4+}$ cations and the hollow sites, respectively. The fat band representation in Fig. 4b shows two important features: (i) conduction bands of CrSBr locate in the gap of MoTe₂; (ii) valence and conduction bands of MoTe₂ are not much affected by CrSBr, suggesting a weak hybridization between CrSBr and MoTe₂. The Berry curvature map in the 2D Brillouin zone (Fig. 4c) shows opposite Berry curvatures in the vicinity of $K_+$ and $K_-$ valleys. These illustrate that the coupled spin and valley physics is remained in CrSBr/MoTe₂.

To quantitatively determine the valley splitting in CrSBr/MoTe₂, we adopt an energy scale [35], $\Delta_{val}^{c/v,\tau} = E_\uparrow^{c/v,\tau} - E_\downarrow^{c/v,-\tau}$. Here, $v$ ($c$) denotes valence (conduction) bands; $K_\pm$ are distinguished by index $\tau = \pm$. According to this definition, $\Delta_{val}^{v,+}$, $\Delta_{val}^{v,-}$, $\Delta_{val}^{c,+}$ and $\Delta_{val}^{c,-}$ are estimated as -37.9, -8.8, 8.6 and 8.7 meV, respectively. We sees that $\Delta_{val}^{v,+}$ in CrSBr/MoTe₂ is sizable, corresponding to the valley splitting generated by a magnetic field of 379 Tesla [35]. It is worth noting that this value is much larger than the counterparts in CrI$_3$/WSe$_2$ [16,46,47] and CrI$_3$/MoTe$_2$ [49]. More remarkably, the smallest energy for the band edge vertical optical transition without spin flip in two valleys reaches a giant value of 46.5 meV. Thanks to the time reversal symmetry breaking and the nonvanishing Berry curvature, CrSBr/MoTe$_2$ has a nonzero anomalous hall conductivity, $\sigma_{xy}$, when Fermi level lies between the valence band maxima of K$_+$ and K$_-$ valleys (Fig. 4d). Taking together the anomalous hall conductivity and sizable valley splitting, it is conceivable that a spin- and valley-polarized Hall current can be generated in CrSBr/MoTe$_2$ when applying an in-plane electric field [51], thus providing applications in valleytronics.

**DISCUSSION**

Finally, we find that CrTe$X$ ($X$=Cl, Br, I) MLs can host magnetic skyrmion states in an external magnetic field, because of the strong SOC of Te and the symmetry reduction. This is a very attractive feature for diverse applications as discussed in the literatures for the studies of other magnetic systems [52-54]. To characterize the presence of magnetic skyrmions in MC simulations, we calculate the topological charge $Q$ which is defined as: [55]

$$Q = \frac{1}{4\pi} \int \mathbf{m} \cdot \left( \frac{\partial \mathbf{m}}{\partial x} \times \frac{\partial \mathbf{m}}{\partial y} \right) dxdy \qquad (3).$$

In Eq. (3), $\mathbf{m}$ is a normalized magnetization vector; $x$ and $y$ are in plane coordinates. On a discrete spin lattice, Eq. (3) is evaluated by summing over the solid angle $\Omega$ of three spins according to the Berg formula [56,57]. Supplementary Fig. 8 shows the topological charge $Q$ of CrTeI as a function of temperature (T) and out-of-plane external magnetic field ($B$). Through examining the spin textures under different T and $B$, we find that the red area with large negative value of $Q$ corresponds to the formation of magnetic skyrmion lattices in CrTeI. Because of the strong DMI, magnetic skyrmion lattices may exist in a large T-$B$ parameter space, with T up to 80 K and $B$ from 1 to 8 Tesla. CrTeCl and CrTeBr also form magnetic skyrmion lattices but in a smaller T-$B$ region (Supplementary Fig. 9). Hence, we recommend CrTeI ML as the most promising 2D platform for the realization of magnetic skyrmions.

In summary, based on systematical first-principles studies on vdW Janus ML Cr$YX$ ($Y$=S, Se, Te; $X$=Cl, Br, I), we find that CrS$X$ ($X$=Cl, Br, I) are useful FM semiconductors with high Curie temperatures up to 176 K and large induced magnetic moments on the ligand $S^{2-}$ anions. Remarkably, the long-sought QAHE with a large gap of 19 meV and a sizable valley splitting of 37.9 meV are achieved through the magnetic proximity effect in vdW heterostructures CrSBr/Bi$_2$Se$_3$/CrSBr and MoTe$_2$/CrSBr, respectively. Furthermore, CrTe$X$ ($X$=Cl, Br, I) may host magnetic skyrmion states under external magnetic fields. Our work unveils the promising multifunctionalities of 2D vdW Janus magnet Cr-based dichalcogenide halides and reveals their potential for diverse applications in topotronic and valleytronic devices.

## METHODS

### First-principles calculations

Our first-principles calculations based on the density functional theory (DFT) are performed using the Vienna *Ab initio* Simulation Package with the generalized gradient approximation [58,59]. Core-valence interactions are described by projector-augmented wave pseudopotentials [60,61]. We utilize an energy cutoff of 350 eV for the plane-wave expansion and fully relax lattice constants and atomic positions until the force acting on each atom is smaller than 0.01 eVÅ$^{-1}$. To take into consideration the strong correlation effect among Cr $3d$ electrons, we adopt $U$=3.0 eV and $J_H$=0.9 eV [15]. As discussed in Supplementary Note 8, we obtain similar results when different U values are employed. In building vdW heterostructures, we use an inplane lattice constant $a_{BS}$=4.16 Å of the relaxed bulk BS for CrSBr/BS/CrSBr and $a_{MT}$=3.55 Å of MoTe$_2$ (MT) ML [62] for MoTe$_2$/CrSBr. When relaxing structures, the first Brillouin zone is sampled by 12×12×1, 6×6×1, and 12×12×1 Γ-centered Monkhorst–Pack $k$ meshes for Cr$YX$, CrSBr/BS/CrSBr, and MoTe$_2$/CrSBr, respectively. We add a vacuum space of 12 Å between slabs along the normal axis to eliminate the spurious interactions. To obtain the accurate magnetic anisotropy energies of Cr$YX$ that arise from SIA, a very dense Γ-centered Monkhorst–Pack $k$ mesh of 24×24×1 is used to sample the first Brillouin zone and the total energy convergence criterion is set to be 10$^{-7}$ eV. Magnetic anisotropy energies are determined by computing the total energy difference with magnetic moments of Cr$^{3+}$ ions being parallel and perpendicular to the plane of the Cr$YX$ ML. When calculating magnetic anisotropy energies, SOC is explicitly included in self-consistent loops. To correctly describe the weak interaction across the vdW gap in these heterostructures, we employ the nonlocal vdW functional (optB86b-vdW) [63,64]. Berry curvatures and chiral edge states are calculated by the wannier90 [65] and

WannierTools [66].

## DATA AVAILABILITY

All data used in this study are available from the corresponding author upon reasonable request.

## CODE AVAILABILITY

The central codes used in this paper are VASP, Wannier90 and WannierTools. The Detailed information related to the license and user guide for these codes are available at https://www.vasp.at, http://www.wannier.org and http://www.wanniertools.com.

## ACKNOWLEDGMENTS

This project is supported by NSFC-12104518, NKRDPC-2017YFA0206203, NSFC-92165204 and the Startup Grant of Sun Yat-Sen University (No. 74130-18841290). Work at University of California, Irvine is supported by the U.S. Department of Energy, Office of Science, Basic Energy Sciences, under Award DE-FG02-05ER46237. Calculations are performed at Tianhe-2 and NERSC.

## AUTHOR CONTRIBUTIONS

Y. H. and F. X. contributed equally to this work. Y. H. and R. W. conceived the whole project. Y. H. and L. Q. carried out the DFT calculations. F. X. performed the MC simulations. All authors made contribution to the final version of the manuscript.

## COMPETING INTERESTS

The authors declare no competing interests.

**Correspondence** and requests for materials should be addressed to Yusheng Hou.

**Figure captions**

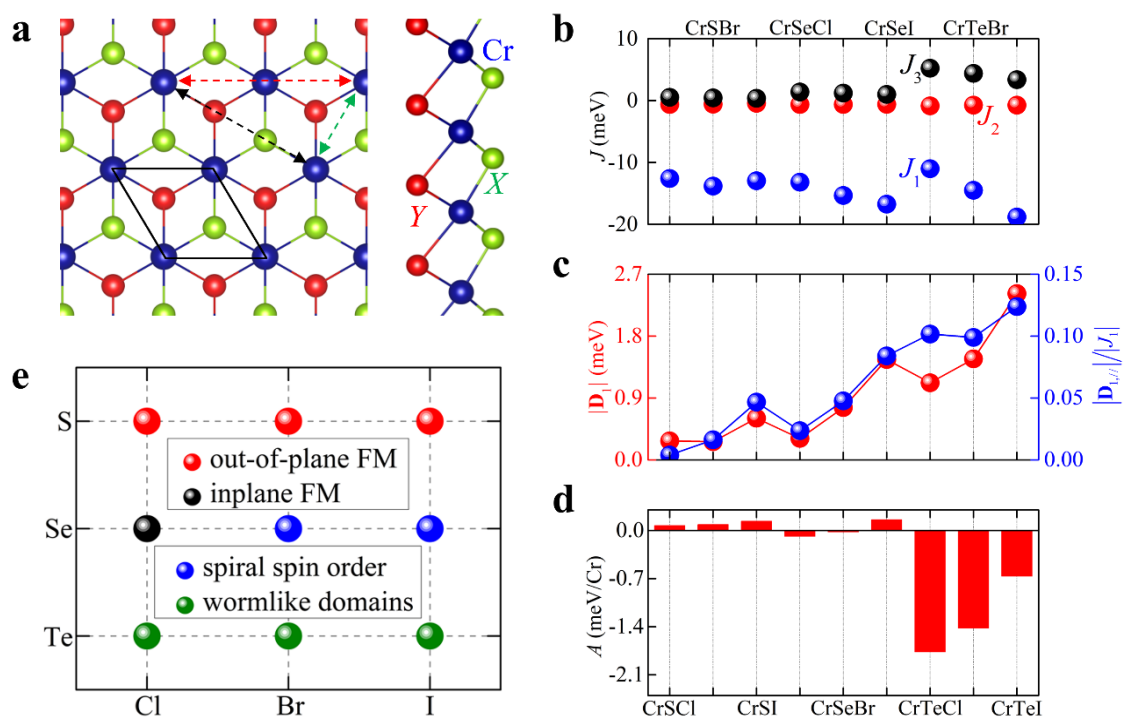

**Fig. 1 Crystal structure and magnetic properties of CrYX. a** Top and side views of the crystal structure. Green, black and red dashed lines show the NN, second-NN and third-NN exchange paths, respectively. **b** Heisenberg exchange parameters $J_i$ ($i$=1,2,3). **c** $|\mathbf{D}_1|$ of the NN DM interaction vector $\mathbf{D}_1$ (red dots) and $\left|\mathbf{D}_{1,//}\right|/|J_1|$ (blue dots). **d** SIA parameter A. **e** MC simulated magnetic ground states of CrYX.

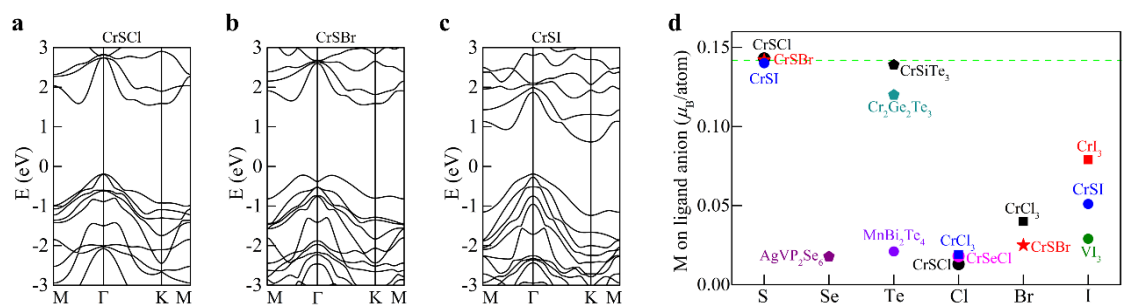

**Fig. 2 DFT+U+SOC calculated band structures** of **a** CrSCl, **b** CrSBr and **c** CrSI. **d** A summary of the DFT calculated induced magnetic moments (M) on the ligand anions of CrSX and other eight representative FM semiconductors. The value of M on $S^{2-}$ ion in CrSBr is highlighted by the green dashed line.

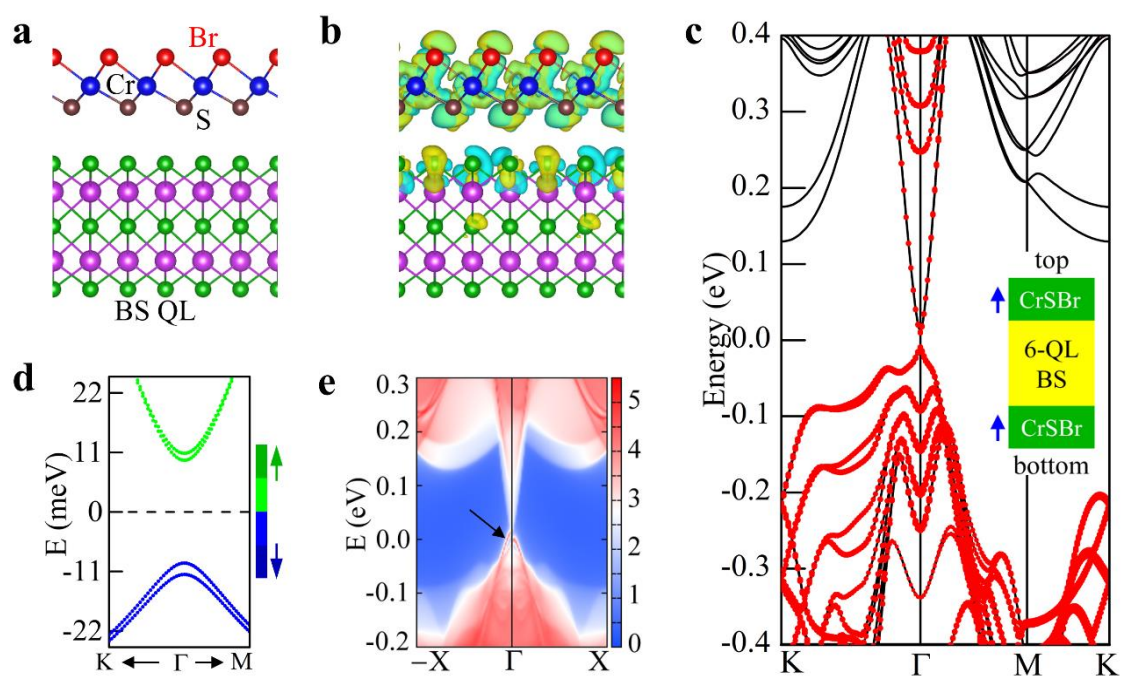

**Fig. 3 Electronic and topological properties of CrSBr/BS/CrSBr. a** Side view of the stable stacking configuration at the CrSBr/BS interface. **b** The spin polarization at the CrSBr/BS interface. Spin-up and spin-down densities are indicated by the yellow and cyan isosurfaces, respectively. The isovalue surface level of spin density is $6 \times 10^{-6}$ eÅ$^{-3}$. **c** DFT+U+SOC calculated band structure. The inset shows the sketch of the magnetizations (represented by blue arrows) of CrSBr MLs. **d** The four bands near the Fermi level. (e) The band structure and the chiral edge state (highlighted by the black arrow) of the one-dimensional CrSBr/BS/CrSBr nanoribbon.

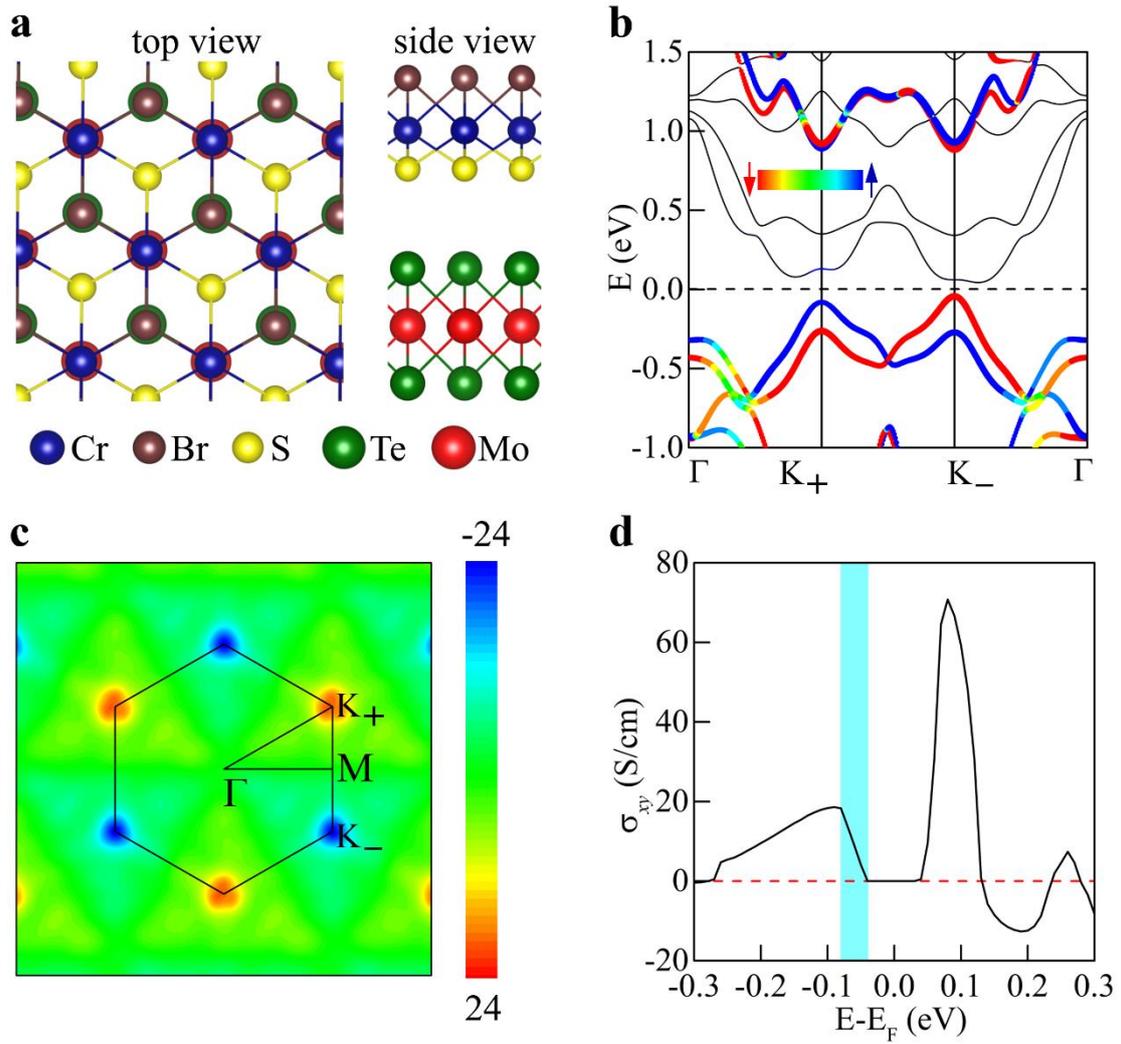

**Fig. 4 Electronic and topological properties of CrSBr/MoTe₂. a** Top and side views

of the stable stacking configuration. **b** MoTe₂-projected (colored lines) band structure.

Color bar indicates the spin projections. **c** Distribution of Berry curvature (in unit of Å²)

over 2D Brillouin zone. **d** Dependence of $\sigma_{xy}$ on the Fermi level. The shaded area denotes the energy window between the two valence-band valley extrema.